\documentclass[sigconf]{acmart}

\AtBeginDocument{%
  }

\copyrightyear{2026}
\acmYear{2026}
\setcopyright{cc}
\setcctype{by-nc-nd}
\acmConference[SIGIR '26]{Proceedings of the 49th International ACM SIGIR Conference on Research and Development in Information Retrieval}{July 20--24, 2026}{Melbourne, VIC, Australia}
\acmBooktitle{Proceedings of the 49th International ACM SIGIR Conference on Research and Development in Information Retrieval (SIGIR '26), July 20--24, 2026, Melbourne, VIC, Australia}
\acmDOI{10.1145/3805712.3808411}
\acmISBN{979-8-4007-2599-9/2026/07}
\begin{document}

\title{Uniboost: Global Coordination with Value Alignment for Fair and Efficient Traffic Allocation}

\author{Ge Fan}
\email{ge.fan@outlook.com}
\orcid{0000-0001-5653-1626}
\affiliation{%
  \institution{Taobao  \& Tmall Group of Alibaba}
  \city{Hangzhou}
  \country{China}}

  \author{Nan Zhao}
  \email{yueqi.zn@taobao.com}
\orcid{0009-0004-7279-2163}
\affiliation{%
 \institution{Taobao  \& Tmall Group of Alibaba}
  \city{Hangzhou}
  \country{China}}

  \author{Kai Meng}
  \email{mengkai.meng@taobao.com}
\orcid{0009-0001-1725-0597}
\affiliation{%
 \institution{Taobao  \& Tmall Group of Alibaba}
  \city{Hangzhou}
  \country{China}}

\author{Cong Luo}
\email{hengying.lc@taobao.com}
\orcid{0009-0004-4941-106X}
\affiliation{%
 \institution{Taobao  \& Tmall Group of Alibaba}
  \city{Hangzhou}
  \country{China}}

    \author{Yang Fu}
    \email{yiang.fy@taobao.com}
\orcid{0009-0002-0403-0832}
\affiliation{%
 \institution{Taobao  \& Tmall Group of Alibaba}
  \city{Hangzhou}
  \country{China}}

\author{Huiping Chu}
\authornote{Corresponding Author.}
\email{yueqi.chp@taobao.com}
\orcid{0009-0003-1359-8670}
\affiliation{%
 \institution{Taobao  \& Tmall Group of Alibaba}
  \city{Hangzhou}
  \country{China}}

\author{Jialin Liu}
\email{miao.ljl@taobao.com}
\orcid{0009-0007-5787-5202}
\affiliation{%
 \institution{Taobao  \& Tmall Group of Alibaba}
  \city{Hangzhou}
  \country{China}}

\author{Yuning Jiang}
\authornotemark[1]
\email{mengzhu.jyn@taobao.com}
\orcid{0000-0003-1665-3025}
\affiliation{%
  \institution{Taobao  \& Tmall Group of Alibaba}
  \city{Beijing}
  \country{China}}

\author{Bo Zheng}
\email{bozheng@alibaba-inc.com}
\orcid{0000-0002-4037-6315}
\affiliation{%
  \institution{Taobao  \& Tmall Group of Alibaba}
  \city{Beijing}
  \country{China}}

\renewcommand{\shortauthors}{Ge Fan et al.}

\begin{abstract}
With the rapid evolution of internet services, recommendation systems have become indispensable. In particular, the blending (re-ranking) stage plays a pivotal role in allocating traffic across diverse business objectives. However, existing approaches often suffer from coupled allocation plans, score inflation, and a lack of interpretability. To address these challenges, we propose Uniboost, a unified traffic allocation framework. Uniboost introduces a posterior value alignment mechanism that calibrates abstract model scores to anchor metrics with explicit business semantics, significantly enhancing interpretability. Furthermore, it employs an independent linear boosting paradigm to decouple complex weighting schemes, enabling precise attribution of each plan's contribution. We validate the effectiveness of Uniboost through online A/B tests and in-depth data analysis, demonstrating three key findings: 1) Reducing the overall weight of weighted scores effectively mitigates unintended business interference, yielding a more efficient micro-level traffic allocation strategy; 2) Post-hoc analyses and aggregated dashboards provide intuitive, macro-level insights that guide the design of the overall traffic allocation mechanism; 3) The proposed “Effective Completion Score” serves as an easily obtainable post-metric that offers a reliable anchor for content recommendation pipelines. Collectively, our experiments show that Uniboost not only improves traffic allocation efficiency and recommendation performance at the micro level but also provides macro-level guidance for system iteration. Thus, this work provides an efficient and controllable traffic regulation solution for large-scale industrial recommendation systems.
\end{abstract}

\ccsdesc[500]{Information systems~Information retrieval}
\ccsdesc[500]{Information systems~Data mining}
\ccsdesc[500]{Information systems~Recommender systems}
\ccsdesc[300]{Information systems~Summarization}
\ccsdesc[300]{Human-centered computing~Interactive systems and tools}

\keywords{Traffic Allocation, Recommendation System, Value Alignment}

\maketitle

\section{Introduction}
\begin{figure*}[htbp]
\centering
\includegraphics[width=1\textwidth]{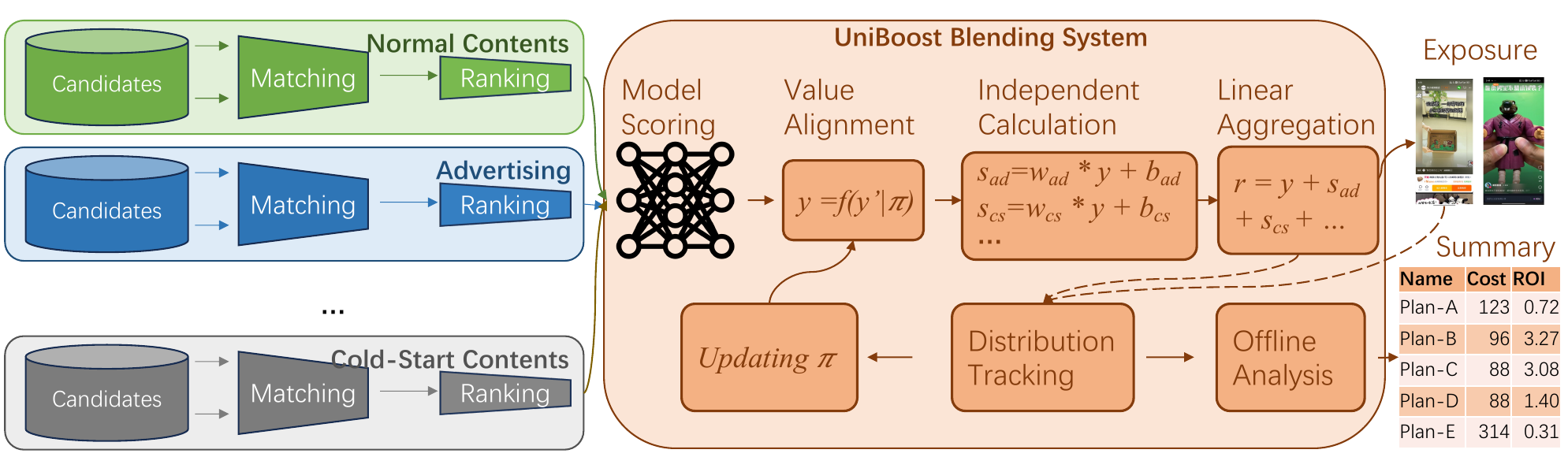}
\caption{Overview of Uniboost Blending System. The figure illustrates the boosting process for two example content types: advertisements (ad) and cold-start (cs) contents. When $w_{ad}=0$
and $b_{ad}$ is set to the output score from the PID-based ad delivery controller, the original PID regulation mechanism for ads is preserved. Conversely, when a non-zero weight $w_{cs}$ is assigned and  $b_{cs}=0$ the formulation reduces to the conventional Boosting approach used for cold-start content.}
\label{fig:framework}
\end{figure*}

With the rapid advancement of internet technologies, search and recommendation systems have become integral to online applications, driving user engagement and business growth. As the core engine of these systems, recommendation algorithms have undergone significant evolution. Traditional systems relied on collaborative filtering to recommend items based on user behavior similarity \cite{linden2003amazon, fan2022pppne, chen2023topic, chen2023neural}. Today, however, industrial-grade recommendation pipelines have evolved into complex multi-stage architectures \cite{ma2022online, wang2022hybrid, fan2022field,  zhao2025multi, movin2025zero}, typically comprising Matching, Ranking, and Blending (Re-ranking) stages, as illustrated in Figure \ref{fig:framework}. The Matching stage retrieves a broad set of candidate items from a massive pool. The Ranking stage employs sophisticated models to score these candidates precisely, filtering for the most relevant content. Finally, the Blending stage merges candidates from diverse sources and applies traffic allocation mechanisms to select the final content for user exposure.

In practical industrial systems, different content types (e.g., ads vs. organic videos) often operate on independent pipelines due to variations in data distribution and business objectives. For instance, the Ranking system for organic video prioritizes user watch time, whereas the Ranking system for ads focuses on ad revenue \cite{chen2022meta, tran2023attention, puthenputhussery2025large, busolin2025efficient, liu2025multi, ranganathan2025zero}. Since ads typically exhibit lower posterior business metrics (e.g., completion rate) compared to organic videos, ranking solely by organic metrics would severely limit ad exposure, reducing commercial revenue \cite{yan2020ads, li2024deep}. To balance user experience with commercial goals, the Blending system often employs guaranteed delivery mechanisms. Specifically, regulation algorithms (e.g., PID controllers \cite{ang2005pid,jambor2012using}) calculate a regulation score based on historical ad exposure rates and add it to the ad blending score to boost visibility \cite{zhang2016feedback, karlsson2018control}. Furthermore, to maintain a healthy ecosystem and mitigate the Matthew Effect \cite{fan2022mv, wang2024preference, xu2024optimizing}, the system must allocate traffic to cold-start content and explore user potential interests. This is typically implemented during the blending stage through boost weighting, where a scaling factor $w$ is applied to the original blending score to increase the exposure likelihood of such content.

However, the proliferation of operational campaigns (e.g., promotions, alliances) introduces a series of short-term traffic allocation plans. In traditional pipelines, these plans accumulated weights cause blending scores to inflate, losing their physical semantics and making them difficult to interpret. Besides,  multiple plans are tightly coupled, interfering with each other and hindering the iteration and optimization of the Blending system. While recent works have attempted to address similar issues in ad systems by aligning user value with commercial value \cite{chen2022hierarchically}, these solutions cannot be directly applied to content recommendation systems due to the following challenges:

\begin{enumerate}
\item \textbf{Posterior Target Selection}. Unlike ad systems where the target (revenue) is stable, content recommendation involves complex, multi-faceted goals (e.g., watch time, clicks, comments). The impact of these targets on the Boost framework remains under-explored, complicating the selection of a robust posterior target.

\item \textbf{Unified Weighting Framework}. Existing weighting schemes are fragmented; ad guaranteed delivery uses PID methods, while cold-start exploration uses Boost weighting. Designing a unified framework that accommodates these diverse mechanisms with minimal disruption to existing business logic is non-trivial.

\item \textbf{Attribution of Weighting Costs}. Due to the coupling of weighting plans, their effects overlap. Accurately evaluating and tracking the cost and contribution of each individual plan remains a significant system design challenge.

\end{enumerate}

To address these challenges, we propose Uniboost, a novel traffic allocation framework that unifies weighting and regulation for diverse content types within the Blending system.

First, through correlation analysis with various posterior targets, we identify "Effective Completion Rate" as our Anchor Metric. By aligning the distribution of blending scores with this completion rate, we transform abstract scores into values with real-world business semantics without altering the ranking order.
Second, based on these aligned scores, we design a Unified Boosting Paradigm with Bias. This paradigm not only covers existing weighting schemes but also ensures all weighted scores retain physical interpretability. Uniboost calculates independent boosting gains for each business plan and aggregates them via linear summation. This enables precise attribution, allowing us to quantify the contribution of each scheme to the final ranking.

In deployment, the online serving stage sorts and outputs content based on these weighted scores to improve micro-level performance. The near-line stage collects weighted and raw scores to update distribution alignment parameters. The offline stage aggregates data to provide macro-level guidance for system iteration.

In summary, our contributions are as follows:

\begin{enumerate}
\item We propose Uniboost, a unified weighting framework that imbues complex blending scores with business value through posterior value alignment. It decouples complex online weighting schemes, providing reliable quantitative evaluation for each individual plan.

\item We design an independent linear boosting paradigm that enables precise attribution of traffic allocation costs. This resolves the score inflation issue and significantly enhances the interpretability of the blending mechanism.

\item  We deploy Uniboost in a large-scale recommendation system via A/B testing. Results demonstrate that our approach not only improves traffic allocation efficiency at the micro level but also provides actionable guidance for system iteration at the macro level.

\end{enumerate}

\section{Proposed Method}

This section details the architecture of our proposed Uniboost system. As illustrated in Figure \ref{fig:framework}, Uniboost is deployed at the Blending stage of the recommendation system. Its primary objective is to support flexible business intervention strategies while maintaining the overall efficiency of the recommendation pipeline. The framework comprises two core components: the Online Serving Pipeline and the Near-line/Offline Statistics System, forming a closed-loop mechanism from strategy execution to performance feedback.
\subsection{Online Serving Pipeline}
The online serving process constitutes the core of Uniboost, designed to complete the blending of candidate items within millisecond-level latency. The workflow consists of four key stages.
\subsubsection{Model Scoring}
First, the system aggregates outputs from upstream retrieval and blending stages to form the final candidate set $\mathcal{V}$, This set is fed into the blending model to generate a raw prediction score $y'_{v}$ for each candidate item$v \in \mathcal{V}$. This score reflects the model's estimation of the user's overall preference.

\subsubsection{Value Alignment}
Raw model scores  $y'_{v}$ often lack explicit business semantics, making direct weighted intervention uncontrollable. To address this, we introduce a Value Alignment Module. This module maps the model scores to a specific posterior target space (defined as the "Anchor Target," e.g., "Effective Completion Rate") by utilizing global alignment parameters
$\pi$ The alignment is formulated as:
\begin{align}
y_v = F(y'_v |\pi)  = \frac{y'_v \mu_{anchor}}{\mu_{score} }
\end{align}
where $\mu_{score}$ and $\mu_{anchor}$ denote the global means of the raw blending scores and the anchor target scores, respectively.

This step calibrates abstract model scores into Expected Values with clear business implications. This ensures that subsequent intervention weights possess interpretable physical significance, bridging the gap between model optimization and business goals.

\subsubsection{Independent Boosting Calculation}
Within the unified value space, the system calculates independent weighted gains for distinct business plans. For any given plan $p \in  \mathcal{S}_{p}$, the boost score $s_{p,v}$ for item $v$ is computed as
\begin{align}
s_{p,v} = \mathbf{1}_{p}(v) * (w_{p}* y_v + b_{p})
\end{align}
where $\mathbf{1}_{p}(v)$ is the indicator function, taking the value 1 if item $v$ belongs to the plan and 0 otherwise;
$w_p$ and $b_p$ are learnable hyperparameters representing the weight coefficient and bias term, respectively.

Since $y_v$ is aligned to the anchor metric, the calculated boost score $s_{p,v} $ shares the same dimensionality as the anchor target. This implies that all business interventions are quantified as specific contributions to the anchor metric, achieving standardized measurement of strategy effects.

\subsubsection{Linear Aggregation}
Finally, the system fuses the base value and all business plan gains via linear summation to obtain the final ranking score $r_v$
\begin{align}
r_v = y_v + \sum_{p \in S_{plan}}{s_{p,v}}
\end{align}
where $S_{plan}$ represents the set of all configured weighting schemes online.

We employ linear summation rather than non-linear fusion to guarantee Additivity. This allows for precise Attribution, enabling us to quantify the contribution of specific schemes to the final ranking result. This feature significantly facilitates downstream strategy tuning and optimization.

Ultimately, the candidate set $\mathcal{V}$ is sorted in descending order based on $r_v$, truncated, and exposed to the user.
\subsection{Near-line/Offline Monitoring System}

To ensure system stability and support iterative strategy optimization, Uniboost establishes a comprehensive data statistics and feedback mechanism, consisting of two main phases.
\subsubsection{Distribution Tracking \& Caching}
This module tracks and caches data distributions across key pipeline stages in real-time, including raw blending scores $y'_v$, aligned value scores $y_v$ and weighted boost scores $s_{p,v}$ for each plan. These fine-grained data points provide critical support for monitoring Distribution Shifts and diagnosing online anomalies.

\subsubsection{Aggregation \& ROI Analysis}

This module aggregates tracking data for downstream strategy evaluation. Specifically, the system statistics the overall resource consumption (e.g., traffic share) and business performance (e.g., improvement in the anchor metric) for each plan. Based on this, we calculate the Return on Investment (ROI) for each weighting plan. For example, suppose we have a plan $p$ aimed at increasing exposure Visit View (VV). Then, the ROI for this plan can be defined as:
\begin{align}
\mathrm{ROI}^{VV} = \frac{\Delta^{VV}_p}{\mathrm{Cost}_{p}}
\end{align}

This metric directly reflects the deployment efficiency of each plan, guiding operators to dynamically adjust parameters. Consequently, this enables globally optimal resource allocation based on data-driven insights.

\section{Experiments and Discussion}
To comprehensively evaluate the impact of the proposed Uniboost framework on the end-to-end recommendation pipeline, we conducted large-scale online A/B testing to validate its effectiveness from both micro and macro perspectives. Specifically, we established a control group using the original weighted pipeline, in which the scores alignment is performed only after weighting, and a treatment group implemented with the unified Uniboost weighting scheme. We selected the Effective Completion Rate as the anchor metric for content recommendation. Additionally, we performed an in-depth online data analysis to justify the selection of this anchor metric. Overall, this section addresses the following three Research Questions (RQs):

\begin{enumerate}
\item [RQ1] Can Uniboost outperform baseline schemes by providing a more efficient traffic allocation strategy at the micro level, thereby enhancing recommendation performance?
\item [RQ2] Can the proposed system provide macro-level guidance for the overall traffic allocation mechanism through the analysis of individual weighting schemes?

\item [RQ3] Is "Effective Completion Rate" a robust choice for the anchor metric in this framework?

\end{enumerate}

\begin{figure*}[htb]
\centering
\includegraphics[width=1\textwidth]{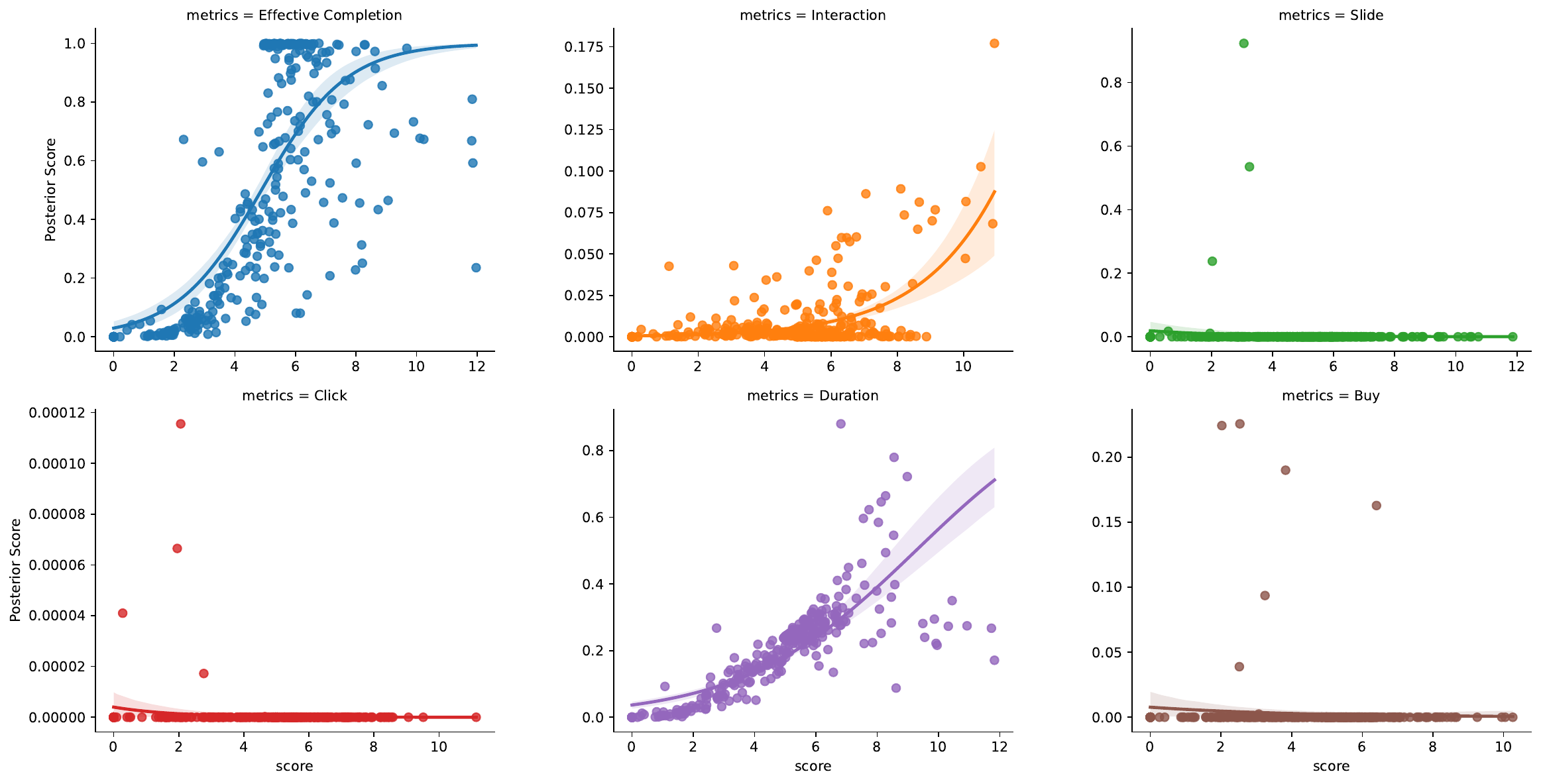}
\caption{Posterior Business Metrics vs. Original Blending Score $y'$.}
\label{fig:metrics}
\end{figure*}

\subsection{Overall Performance}
The results of the online A/B experiments are summarized in Table \ref{tab:result_all}. The key metrics are defined as follows: Visit View (VV) represents the total number of content impressions; Valued VV refers to the count of contents played for more than 3 seconds; Duration indicates the total consumption time; and Valued Score is a core composite metric of the scenario, constructed from diverse user behaviors such as clicks and comments, used to evaluate the overall user experience.

\begin{table}[tbp]
\caption{Relative Improvement in Online A/B Test.}
\centering
\begin{tabular}{l|l|l|l|l}
\hline
    Metrics & VV & Valued VV  & Duration   & Valued Score  \\
    \hline
    \hline
    Overall & +1.69\% & +3.07\% & +0.65\% & +2.54\%  \\ \hline
\end{tabular}
\label{tab:result_all}
\end{table}

The results demonstrate that the proposed method achieved consistent improvements across all core metrics, confirming the overall system-level benefit of Uniboost. Monitoring data reveals that compared to the Control Group, the Treatment Group achieved a 92.21\% reduction in ad weighting scores and a 95.81\% reduction in organic video weighting scores, yet maintained stable exposure shares for all weighted content categories. This indicates that Uniboost effectively mitigates the weight explosion issue prevalent in the original system. By significantly reducing the interference of regulation strategies on core business objectives while guaranteeing required exposure levels, Uniboost enhances the overall performance of the recommendation system. These results corroborate Uniboost's capability to decouple inter-dependencies among various boosting plans, addressing the limitations of traditional boosting. Consequently, the proposed scheme enables more efficient traffic allocation at the micro level, significantly enhancing the overall efficiency of the recommendation pipeline.

\subsection{Macro-level Guidance via Offline Analysis}
To verify whether the proposed scheme can provide macro-level guidance for traffic mechanism regulation, we statistically analyzed the Cost, VV Lift, and ROI of each business plan. Based on this analysis, we identified Plan-A as having the lowest ROI. We subsequently constructed an ablation study by removing Plan-A from the online configuration. The experimental results are presented in Table \ref{tab:result_off}.

\begin{table}[bp]
\caption{Relative Improvement of removing Plan-A.}
\centering
\begin{tabular}{l|l|l|l|l}
\hline
    Metrics & VV & Valued VV  & Duration   & Valued Score  \\
    \hline
    \hline
    w/o Plan-A & +0.95\% & +2.83\% & +3.49\% & +4.13\%  \\ \hline
\end{tabular}
\label{tab:result_off}
\end{table}

As observed, removing Plan-A led to a significant improvement in core online metrics.
Overall, exposure VV shows a slight increase, while Valued VV, Duration, and Valued Score significantly improve, indicating that the content recommended by the new pipeline is indeed well-received by users. This indicates that Plan-A indeed incurred high costs for marginal gains, thereby negatively impacting the overall performance of the recommendation pipeline. These results corroborate that the analytical reports generated by Uniboost not only track the costs and benefits of individual schemes but also provide actionable guidance for macro-level traffic allocation. Consequently, this enhances the overall efficiency of the recommendation chain and accelerates the iteration speed of strategy optimization.

\subsection{Selection of the Anchor Metric}

We analyzed the correlation between the blending model prediction scores and common post-metrics, including Effective Completion, Play Duration, Click, Buy, Interaction, and Slide. Figure \ref{fig:metrics} illustrates the distribution differences of various post-metrics under different prediction scores, where the x-axis represents the model prediction score and the y-axis represents the probability of the post-metric.

We observed that Buy, Interaction, and Slide are unsuitable as anchor metrics due to their extreme sparsity and high volatility across prediction scores. While Interaction and Play Duration are relatively denser, their distributions diverge from the score distribution, particularly in the long-tail region (i.e., high score values), where the prediction error increases significantly. In contrast, the Effective Completion metric remains relatively stable. Its values are uniformly distributed across various intervals of the blending scores, and the calibration error remains low and stable across all intervals. This validates that in video recommendation feeds, Effective Completion serves as a robust anchor data point for aligning the value of blending scores.

\section{Conclusion}
This work focus on the blending stage within the recommendation system pipeline. In this paper, we propose Uniboost, a unified traffic allocation framework grounded in posterior value alignment. Uniboost tackles the challenge of interpretability in traditional re-ranking by identifying stable posterior targets. It aligns abstract model output scores—which often lack physical semantics—with anchor metrics that possess real-world business value. This alignment significantly enhances the transparency and controllability of the traffic allocation mechanism. Building on this foundation, Uniboost replaces the original complex, coupled, and diverse boosting processes with an independent linear boosting scheme. This innovation provides a more efficient and streamlined solution for overall traffic regulation, decoupling inter-dependent strategies for clearer optimization.

Extensive experiments and analyses demonstrate the dual benefits of our approach. At the micro level, Uniboost allocates traffic more efficiently, improving the operational performance of the overall pipeline. At the macro level, it provides strategic guidance for traffic mechanism design through long-term tracking and analysis of weighting costs for each scheme. Consequently, Uniboost not only optimizes immediate recommendation performance but also supports the sustainable iteration and evolution of the entire recommendation system. This solution has been fully deployed in the Taobao Content Feeds recommendation system and is stably operating as the core traffic allocation system.

\section*{Acknowledgment}

We are deeply grateful to Zhibo Xiao, Dimin Wang and Jialin Zhu for their generous sharing of insights on value alignment in item recommendation systems. Their thought-provoking discussions and practical perspectives, developed through internal collaborative work, have greatly inspired our approach in this study. We sincerely thank all the reviewers and the area chairs for their insightful comments and constructive suggestions, which have greatly improved the quality of this paper. We are also deeply grateful to all collaborators involved in this project, including, but not limited to, the Server Engineering team, Data Science team, Product team, and Operations team, whose support was instrumental in enabling the successful execution of this research.

The authors acknowledge the peoples of the Woi Wurrung and Boon Wurrung language groups of the eastern Kulin Nation on whose unceded lands ACM SIGIR 2026 was hosted. We pay our respects to their Elders past and present, and extend that respect to all Aboriginal and Torres Strait Islander peoples today and their continuing connection to land, sea, sky, and community

\bibliographystyle{ACM-Reference-Format}
\balance
\bibliography{refer}

\end{document}